\documentclass[openacc]{rstransa}

\usepackage{amsmath,amsthm,amsfonts,amssymb}
\usepackage{graphicx}
\usepackage{color}
\usepackage[normalem]{ulem}
\definecolor{sepia}{RGB}{112,66,20}
\newcommand{\kbT}{k_{\scriptscriptstyle B}T}

\begin{document}
\title{The Physics of Stratum Corneum Lipid Membranes}

\author{
Chinmay Das$^{1}$ and Peter D. Olmsted$^{2}$}

\address{$^{1}$School of Mathematics, University of Leeds, Leeds LS2 9JT, UK\\
$^{2}$Department of Physics and Institute for Soft Matter Synthesis \&  Metrology, Georgetown University, Washington DC, 20057, USA}

\subject{computational physics, soft matter physics, biophysics, physical chemistry}


\corres{Peter D. Olmsted\\
\email{peter.olmsted@georgetown.edu}}

\begin{abstract}
The Stratum Corneum (SC), the outermost layer of skin, comprises rigid corneocytes (keratin filled dead cells) in a specialized 
lipid matrix. The continuous lipid matrix provides the main barrier against uncontrolled water loss and invasion of external pathogens.
Unlike all other biological lipid membranes (like intracellular organelles and plasma membranes), molecules in SC lipid matrix show
small hydrophilic group and large variability in the length of the alkyl tails and in the numbers and positions of groups that 
are capable of forming hydrogen bonds. Molecular simulations provide a route for systematically probing the effects of each
of these differences separately. In this article we present results from atomistic molecular dynamics of selected
lipid bilayers and multilayers to probe the effect of these polydispersities. We address the nature of the tail packing in the gel-like phase, the hydrogen bond network among head groups, the bending moduli expected for leaflets comprising SC lipids, and the conformation of very long ceramide lipids (EOS) in multibilayer lipid assemblies.

\end{abstract}

\maketitle

\section{Introduction and Review}
\begin{figure}[!h]
\vspace*{-7pt}
\centerline{\includegraphics[width=0.9\linewidth]{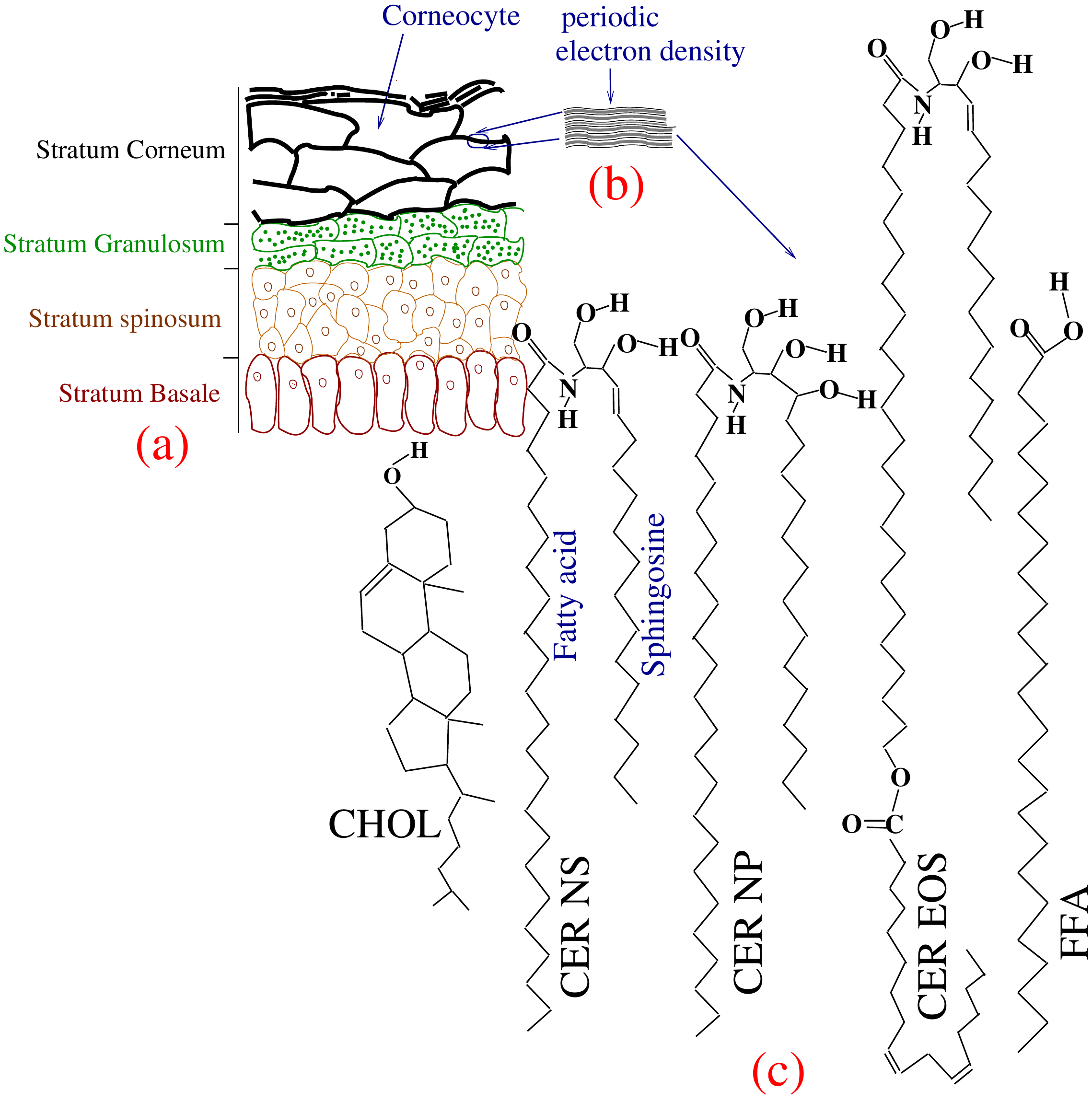}}
\caption{Schematic representation of (a) the layers in the epidermis. The stratum corneum (SC), the topmost layer of epidermis, contains rigid corneocytes. (b) Sketch of the electron density profile in the lipid
matrix between corneocytes, showing the lamellar structure \cite{norlen2003cryotransmission}. (c) Schematic structure of some of the SC lipids, viz.: CHOL, CER NS 24:0, CER NP 24:0, CER EOS 30:0, and FFA 24:0.
For CER NS, the two tails are marked to show the sphingosine (S-chain) and the fatty acid (N-chain) chains. 
The N-chain of CER EOS drawn here has a 30 carbon saturated fatty acid linked to a linoleic acid.  }
\label{fig.intro}
\vspace*{-5pt}
\end{figure}

The outer layer of skin, the epidermis, presents itself as a collection of distinct layers in histological sections (Fig.\ref{fig.intro}a). The stratified
structure is maintained by cell divisions at the innermost part of the epidermis, overproduction of specialized lipids that are
transported to the extracellular space, overproduction of keratin proteins, followed by cell death and eventual desquamation at the 
outermost part. The stratum corneum (SC), the outermost $\sim 20\,\mu\textrm{m}$ section of epidermis, is often qualitatively
described as a brick and mortar structure \cite{michaels.sc.brick.75}.
In this picture corneocytes, flattened disc-like keratin network-filled cells lacking 
all cytoplasmic organelles, take the role of the bricks and a multilayer structure of specialized lipids forms the mortar phase. 
The continuous lipid matrix in SC provides the main barrier against water loss and against invasion of external chemicals and 
pathogens \cite{elias.sc.rev.05}. 

Ceramides (CER) form the major two-tailed component of the SC lipid matrix.  CER molecules contain a sphingosine tail (`S'-chain)
linked to a fatty acid tail (`N'-chain) via an amide bond. Skin ceramides differ from other ceramides  by having an N-chain that is  highly 
polydisperse in length and typically significantly longer than the sphingosine chain \cite{farwanah.sccomp.cer.05}. In addition to  
length polydispersity, SC ceramides show additional polydispersity in the numbers and placements of hydrogen-bond-capable hydroxyl groups
in the head region. In some cases  a doubly unsaturated linoleic acid is conjugated to the N-chain of CER. 
This large polydispersity leads to more than 300 distinct CER molecules  in the SC. The usual
nomenclature for CER includes the nature of the fatty acid motif followed by that of the sphingosine motif. When required, the
number of carbons in the fatty acid motif is included as a postfix. Thus, CER NS 24:0 (Fig.\ref{fig.intro}c) refers to ceramide with
the fatty acid motif having a 24 carbon fully saturated {\em non-hydroxyl} fatty acid linked to an unmodified
sphingosine. The notation 24:0 refers to a acyl tail of length 24 carbons with 0 unsaturated carbon-carbon bonds. The sphingosine motif for CER NP is a {\em phytosphingosine} (with an additional OH group at the
4th carbon position as compared to CER NS). 
Most ceramides in the SC have tails with lengths of 
16 (sphingosine tail or S-chain; 2 carbon atoms of 18 carbon sphingosine resides in the head group) and 22-32 (fatty acid tail or N-chain). 
There are also ceramides whose
N-chains are further linked to a linoleic acid giving rise to exceptionally long N-chain. One example of such long-chained ceramide is CER EOS.
Without accounting for the linked linoleic acid, the N-chains of CER EOS typically have 30 to 34 carbon atoms.
From the sequence of appearance during gel-permeation experiments, CER EOS, CER NS, and CER NP have also 
been often referred respectively as CER1, CER2, and CER3.

Free fatty acids (FFA) with similar length polydispersity  \cite{norlen.sccomp.ffa.98}, 
and cholesterol (CHOL) form the other two dominant component of SC lipids. We refer to the different FFA by adding the number
of carbon atoms in the molecule. Thus we refer to lignoceric acid as FFA 24:0.
The relative abundance of the components vary between individuals and within the same individual depending on the body site \cite{norlen.sccomp.99}. 

The abundance of saturated long alkyl tails and the lack of polarizable head groups in SC lipids endow them with very different 
properties than plasma membranes.  In pure form any of the CER or FFA molecules are crystalline or solid  below $\sim 80^{\circ}\,\textrm{C}$. In 
a multilayer arrangement SC lipids show limited hydration, and polydisperse SC lipid mixtures remain in a `gel' (glassy) phase
at physiologically relevant temperatures. 

The detailed molecular arrangement in SC lipid matrix is still highly debated: Cryo-Electron Microscopy (cryo-EM) images 
\cite{Madison.JID.87,Amoudi.JID.05} of skin slices provide
the most direct visualisation of the lipid structure. The electron density pattern is often found to have alternate major and minor bands 
between corneocytes, prompting the development of a large number of models aiming to explain such a pattern \cite{bowstra.sc.xray.95,schroter.bpj.09,plasencia.sc.confocal.07,iwai.jid.12,mojumdar2013localization, mojumdar2015phase}. This picture breaks
down at regions where corneocyte walls are further apart (lacunar spaces). Cryo-EM often shows lack of lamellar arrangement in
such lacunar spaces, and more complex lipid structures with a checkerboard pattern consistent with  two or three dimensional periodicity have been reported \cite{Amoudi.JID.05}.

Hydration of SC is not homogeneous: under normal hydration most of the water resides in the corneocytes \cite{bouwstra.jid.03} that
contains hygroscopic molecules collectively termed `natural moisturising factor'.  
Water pockets in the lacunar spaces that leaves the lamellar arrangement between closely apposed corneocyte boundaries
unchanged have been observed during extreme  hydration \cite{warner.jid.03}. It is
believed that a layer of CER are covalently bound (at the CER tail end) to the internal protein network on the surface of corneocytes (corneocyte
bound lipid envelope or CLE) \cite{swartzendruber.CLE}.

A large number of experiments exist 
both with lipids extracted from SC and with different combinations of synthetic SC lipids at different molar ratios: 
the structures inferred from such experiments range from lamellar
structures with different periodicities \cite{bowstra.sc.xray.95,schroter.bpj.09} to microporous structures \cite{plasencia.sc.confocal.07}. 
Periodicities inferred from scattering experiments on {\em in~vitro} SC lipid mixtures are conveniently classed as short period periodicity (SPP $\sim 6\,\textrm{nm}$) 
and long period periodicity (LPP $\sim 13\,\textrm{nm}$). Both CER EOS \cite{bowstra.sc.cer1.98} and CHOL \cite{mojumdar2013localization, mojumdar2015phase}
were found to play an important role in stabilizing the LPP structures. 
Cryo-EM images from \textit{ex vivo} SC sections show both SPP and LPP periodicities \cite{iwai.jid.12}.
These diverse results may reflect the limited mobility of 
lipids in the gel phase. We expect that the mobilities in the confined space between corneocytes will be even more limited, 
and that the morphologies observed \textit{in vivo} depend strongly on specific active and  kinetic pathways determined by the biology. 

With this complex mixture of molecules, atomistic simulations offer a useful way to systematically separate out particular 
components and look for their effects in a well defined geometry. In the last decade, molecular dynamics
simulations have yielded a large number of insights into SC lipids, especially when considered as fully hydrated bilayer configurations.
Thus, the simulated systems differ in several aspects compared to the {\em in vivo} SC lipids: (a) the {\em in vivo} structure, which is essentially a 
water-free lipid multilayer confined between corneocytes, is usually replaced in simulations by a single fully hydrated bilayer; (b) the very complex 
lipid mixture is usually replaced by a few selected lipid molecules; (c) simulations are limited in both the system size (typically less than 
1000 lipid molecules) and the longest simulation time  (of order $\mu s$ or less). 

Keeping these limitations of the simulations in mind
CHOL was found to reduce the alkyl tail order parameter and the bilayer thickness for FFA bilayers \cite{holtje.fachol.01}.
Simulations of symmetric CER NS 16:0 \cite{pandit.cer2.06} showed liquid crystalline arrangement of the tails at 
$368\,\textrm{K}$ and weaker perturbation of the water molecules when compared to sphingomyelin bilayers.
Simulations with asymmetric tailed CER NS 24:0 and DMSO (dimethylsulfoxide) \cite{notman.dmso.07} showed that 
DMSO preferentially forms hydrogen bonds with CER and reduces the area compressiblity. At high enough DMSO concentration the 
bilayer arrangement is destroyed, providing a molecular explanation of the permeation enhancing property of DMSO. 
Potentials of mean force calculated from CER NS 24:0 and DMSO simulations \cite{Notman.BPJ.08} showed that the barrier
for a  pore  with a  radius large enough for water permeation in CER NS is prohibitively large ($> 700 \textrm{k}_{\textrm{B}} \textrm{T}$); 
 DMSO was found to drastically lower the barrier height and form a hydrophilic layer inside the channel, which allows easy access for water molecules.
Simulations of different molar ratios of CER NS 24:0, CHOL, and FFA 24:0 \cite{das.bpj.09} showed that the three
component bilayers have smaller area compressibility and smaller variations in the 
excess lateral pressure profile compared
to bilayers composed of any of the isolated three components alone \cite{das.bpj.09}. Also,
the presence of two different tail lengths (18 carbon sphingosine motif and 24 carbon fatty acid motif) leads
to a sandwich structure with the excess carbons in the longer tails forming a disordered zone inside liquid crystalline
ordered leaflets. The effects of oleic acid on the lipid order, hydrogen bonding  and mobilities were studied in three component  bilayers \cite{Hoopes.JPCB.11}. 
Ref.~\cite{Guo.JCTC.13} compared different force-field models and studied the thermotropic phase transitions of CER NS 16:0 and CER NP 16:0, finding that 
CER NP forms more inter-lipid hydrogen bonds than CER NS, which leads to a higher chain melting temperature.

A series of simulations with a single water molecule constrained to remain at a fixed distance $z$ from the midplane was
used to calculate the local (in $z$)  excess chemical potential and diffusivity of water inside SC lipid bilayers \cite{das.smat.09}. A large
 barrier for penetration ($\sim 15 \textrm{k}_{\textrm{B}} T$ at 300K) was found, which ensures that local diffusivity alone is a poor indicator for 
the time to cross the bilayer. The profiles of excess chemical potential and diffusivity can  be used to calculate  the  permeability of SC lipid bilayers to water,   
which was found to be  approximately five orders of magnitude smaller than
DPPC bilayers. A similar computation on a double bilayer of SC lipids (with no water between the two bilayers) 
showed a barrier against swelling because of the abundant inter-leaflet hydrogen bonds between lipids from the two bilayers \cite{Das.sm.14}.
At molar concentrations similar to the SC a fraction of CHOL molecules can remain in the bilayer midplane \cite{das.bpj.09}, and
the transport of CHOL between the two leaflets (flip-flop) is fast enough (a few microseconds) to be observed \cite{Das.sm.14} in computer simulations.

In typical simulations with a preformed bilayer structure the lamellar phase is artificially stabilised because of the periodic boundary
conditions and small system sizes. For a reasonably large system size and realistic representation of the polydispersities in the
SC lipids, randomly oriented SC lipids with $30\,\textrm{wt\%}$ water (average water content in normal SC) lead to an
inverted-micellar arrangement in timescales available in simulations \cite{das2013lamellar}. Even starting from 
lamellar arrangement leads to inverted phases in situations for which the initial conditions of the simulation allow the formation of inverse phases without drastic changes in
the local molecular structure \cite{das2013lamellar}. This preference for inverted phases can also be inferred from the elastic properties extracted from the excess lateral pressure profile measured in long bilayer simulations. The integral of the first moment of the  lateral pressure profile is related to the intrinsic curvature of an individual leaflet, and  for leaflets containing CER with small head groups simulations find \sout{to} a negative curvature (\textit{e.g.} Table~\ref{tab:moduli}), which implies an inverse phase. A  substrate patterned with CER headgroups, which provides local 
sites for CER-CER hydrogen bonding,  can lead to growth of
lamellar structures \cite{das2013lamellar}, which suggests that the CLE plays a crucial role in maintaining the bilayer motif between corneocytes in the  SC.

While providing useful insights, such large-scale simulations with multiple complexities do not allow for 
understanding the contribution of the separate components. In this paper we report simulation results on
SC lipid bilayers and multilayers with the components chosen so that they include different facets of
the complexity of SC lipids and isolate their effects: 
\begin{enumerate}
\item We compare simulations with either CER NS 24:0 or
CER NS 16:0, in order to understand the role of the different lengths of the sphingosine and fatty acid tails 
of CER. In both cases, the sphingosine motif contains 18 carbons (with two of the carbon atoms in the
head). CER NS 16:0 contains 16 carbons as the linked fatty acid, which leads to a  CER with
equal  tail lengths. CER NS 24:0 contains a 24-carbon fatty acid tail, which leads to 
an asymmetric tail-length configuration representative of SC lipids. 

\item To understand the role of head group polydispersity, we
consider bilayers formed with CER NP 24:0. CER NP is identical to CER NS,  except that  the unsaturated double bond
in the 4,5 position of CER NS is saturated in CER NP, and the resulting  additional -OH group at the 4 position
leads to one extra hydrogen bond donor-acceptor group in CER NP, relative to CER NS. We compare simulations using CER NP 24:0, CER NS 24:0,
and a equimolar mixture of the two, to isolate the effect of the head-group. In principle, CER NS has 3 donor and 4 acceptor groups, 
while CER NP has 4 donor and 5 acceptor groups. 

\item We  study a bilayer
with a 1:2:1 mixture of CER NS 22:0, 24:0, 26:0, to probe the effect of polydispersity in the fatty acid tail
length in a given class of CER. 

\item Finally, we also considered a multilayer that contains representative 
complexity of the SC in terms of lipid components and tail polydispersity of both the CER and FFA.
In particular, we use this system to address the configuration of the long-tailed CER EOS lipids in SC lipid multilayers.
\end{enumerate}

\section{Simulations}
\subsection{Force Fields}
For the results reported in this paper, we carried out extended ensemble molecular dynamics simulations of
fully hydrated bilayer and multilayer systems with several different choices of SC lipids  at constant 
temperature and pressure, using the GROMACS software package \cite{gromacs05}. The lipid interactions were described by united atom
Berger force-field \cite{chiu.ff.95, berger.ff.97} with explicit polar hydrogens. 
Water is modelled  with the SPC potential \cite{spcwater} and a cut-off length of
$1.2\,\textrm{nm}$ was chosen for the Van~der Waals interactions. Long-range electrostatics 
are handled using particle-mesh Ewald summation. For details of the 
partial charges and topologies used, 
the reader is referred to earlier publications \cite{notman.dmso.07,das.bpj.09, das2013lamellar}.
In previous simulations \cite{das.bpj.09} we found that all properties behave smoothly between 300 and 360K for
SC lipids with this choice of potentials. For faster equilibration, simulations in this work were performed at  $T=340\,\textrm{K}$.

\subsection{Preparation of ceramide bilayers}
For all the simulations containing only CER molecules, 128 lipid molecules were considered in an initial hairpin conformation,
arranged in symmetric leaflets in excess water (5250 water molecules). Both CER NS and CER NP show numerous crystalline arrangements in bulk, depending on the temperature and crystallization
conditions \cite{dahlen.cer224.xray.79, raudenkolb2003cer3}. In only one 
experimental condition was CER NS 24:0 found to form a crystal with a hairpin arrangement of the molecules (with both tails opposing the headgroup); all other 
 crystal structures have  extended chain arrangements (with the head group between the extended tails) \cite{dahlen.cer224.xray.79}. However, in a single hydrated bilayer the hairpin
arrangement of the lipids is the only possibility. The starting configuration for CER NS 24:0 was
the already-equilibrated configuration from  \cite{das.bpj.09}. All other CER bilayers 
were derived from this configuration by sequential grafting of the required number of OH groups (to get CER NP 24:0)
and either grafting or removing methyl groups at the fatty acid tail of CER NS (to adjust tail lengths). 
Each  such change was followed by energy minimization, short MD simulations in the NVT ensemble followed by MD
simulations in the NPT ensemble. Each  configuration was evolved for $350\,\textrm{ns}$ and no systematic
average evolution was found in any of the measured quantities beyond the first $50\,\textrm{ns}$, which is thus considered to be the equilibration time. 
All reported averages
are from the last $300\,\textrm{ns}$ of the simulations.

\subsection{Simulations with long EOS ceramides}
The extra-long-tailed ceramides  with esterified linoleic acid comprise about 10 mole\% of the total
ceramide content in SC \cite{weerheim.sccomp.01,farwanah.sccomp.cer.05}. 
These long-tailed ceramides have often
been implicated in the signature of long-period periodicities ($\sim 13\,\textrm{nm}$) observed in 
some scattering  experiments\cite{bouwstra.jlr.98}. In a different model trying to explain the
electron density pattern observed in cryo-EM experiments, the long tailed CER EOS have been assumed to have
an extended chain structure \cite{iwai.jid.12}. 
The N-tail of CER EOS 34 contains 52 carbons (including the 18 carbons from the conjugated linoleic acid). 
If fully stretched, the N-tail of CER EOS will be longer than typical bilayer thickness (which is $\sim5.2$nm). 
Thus, in principle, even in the hairpin conformation a CER EOS molecule can span both leaflets of a bilayer and 
protrude into the neighboring bilayer's leaflet. Neutron diffraction of multilayers containing  CER EOS suggested such
a structure with the N-tail of CER EOS connecting three leaflets \cite{kessner.EBJ.08}.
To  understand  the role of the lipid
polydispersity and examine the conformation adopted by the extra-long chain CER EOS,
we consider an equimolar mixture of CER, FFA and CHOL with realistic polydispersity.

This  realistic polydispersity requires a much larger system size. We use the following 
constituents (Table \ref{tab:BigMixture}):
\begin{itemize}
\item CER EOS with the N-chain (excluding the conjugated linoleic acid)
of lengths 30 carbons
(20 molecules), 32 carbons (52 molecules), 33 carbons (20 molecules), and 34 carbons (40 molecules). 
Each N-chain  is further linked to a linoleic acid (18 carbons with two double bonds at 
the 9 and 12 position). Fully stretched, one tail of these CER EOS can span one and half bilayers. In all cases the S-chain was assumed to be 16 carbons long. 
\item CER NS molecules with the  N-chain  containing 22 carbons (68 molecules), 
24 carbons (132 molecules), 25 carbons (68 molecules), 26 carbons (200 molecules), 28 carbons (132 molecules), 
and 30 carbons (68 molecules).
\item CER NP molecules with the N-chain containing
24 carbons (52 molecules), 26 carbons (92 molecules), 28 carbons (140 molecules), 30 carbons (160 molecules),
and 32 carbons (92 molecules). 
\item Fully saturated FFA with the numbers of carbons being 20 (64 molecules),
22 (136 molecules), 24 (536 molecules), 25 (136 molecules), 26 (304 molecules), 28 (136 molecules) and
30 (24 molecules). 
\item 1332 CHOL molecules. 
\end{itemize}
The molar ratios were chosen such that
the fraction of CER EOS implicitly accounts for other long-tailed ceramides found in the SC. Similarly, CER NS and CER NP implicitly account 
for the other short-tailed ceramides found in the SC. 

\begin{table}[!h]
\caption{Composition of polydisperse simulation with EOS, NS, NP, FFA, and CHOL molecules in a multilayer
arrangement in water. There were a total of 25000 water molecules. 
Shown are the number of molecules with different numbers of carbons in the N-chain (for EOS, NS, NP) or in the 
free fatty acids (FFA). The number of carbons in the N-chain of CER EOS excludes the conjugated linoleic acid.}
\label{tab:BigMixture}
\begin{center}
\begin{tabular}{ccccccccccc|c}
\hline
Carbon Number &20 &22 & 24 & 25 & 26 & 28 & 30 & 32 & 33 & 34 &Total\\
\hline
EOS &0 & 0 & 0 & 0 & 0 & 0 & 20 & 52 & 20 & 40 &132 \\
NS & 0& 68 & 132 & 68 & 200 & 132 & 68 & 0 & 0 & 0 &668\\
NP & 0& 0 & 52 & 0 & 92 & 140 & 160 & 92 & 0 & 0 &536\\
FFA& 64 & 136 & 536 & 136 & 304 & 136 & 24 & 0 & 0 & 0 &1336\\
CHOL & -& -& -& -& -& -& -& -& -& -&  1332\\\hline
Total & 64  & 204  & 720  & 204  & 596  & 408  & 272  & 144  & 20  & 40  & 4004
\end{tabular}
\end{center}
\end{table}

We considered a double bilayer
with equal numbers of each species in each of the four leaflets, which allows us to study whether the 
 conformation adopted by CER EOS can couple two bilayers.
Each leaflet was independently prepared by randomly placing the required number of molecules  in
an expanded box in the bilayer ($x\!-\!y$) plane,  with all ceramides in the hairpin straight-tail configuration pointing along the $z$-direction 
and all the lipid head groups in the same plane (at a given $z$). 
The leaflets were energy-minimized and combined to
form the double bilayer structure with sufficient separation between the leaflets to 
accommodate the very long  CER EOS N-chains. 
Continuous water walls were used at the box boundaries  
in the $z$ direction and periodic boundary 
conditions were used in the $x-y$ direction.
To adopt a dense liquid-like configuration the following protocol was repeated: the structure was compressed  in the $x\!-\!y$ plane by 1\%,  and  the leaflets and the continuous  water walls were moved 
closer to each other along $z$ by $0.001\,\textrm{nm}$, and  
the system was energy minimized.
This protocol ensured that the lipids remain in a 
multilayer arrangement while locally deforming. Once the internal 
pressure reached atmospheric pressure, the continuous
water walls were replaced by a layer of 25000 water molecules and periodic boundary conditions were applied in all three directions. 
The head groups were {\em hydrated} by using a short NVT simulation where the lipid molecules were
frozen. Finally the constraints were removed and the system was evolved in the NPT ensemble for $300\,\textrm{ns}$. 
For this simulation we use a group-based cut-off for the electrostatics with a cut-off length of $1.2\,\textrm{nm}$.
At the physiologically relevant pH the SC lipids considered here remain uncharged and neglecting the
long-range nature of the electrostatics interactions makes little difference \cite{das.bpj.09}.
In earlier simulations on the same lipid composition but with  fewer  (20000) water molecules these multilayers were unstable to forming an inverted structure 
\cite{das2013lamellar}. 
The current system with more water molecules remains in a multilayer arrangement for the
entire simulation ($300\,\textrm{nm}$).  Stable layers may arise because  the system with a thicker water layer requires a larger fluctuation in order to nucleate a bent bilayer and explore curved configurations.

\section{Results}

\subsection{Effect of asymmetric CER tail lengths}
\begin{figure}[htbp]
\vspace*{-7pt}
\centerline{%
 \includegraphics[width=0.46\linewidth,clip=]{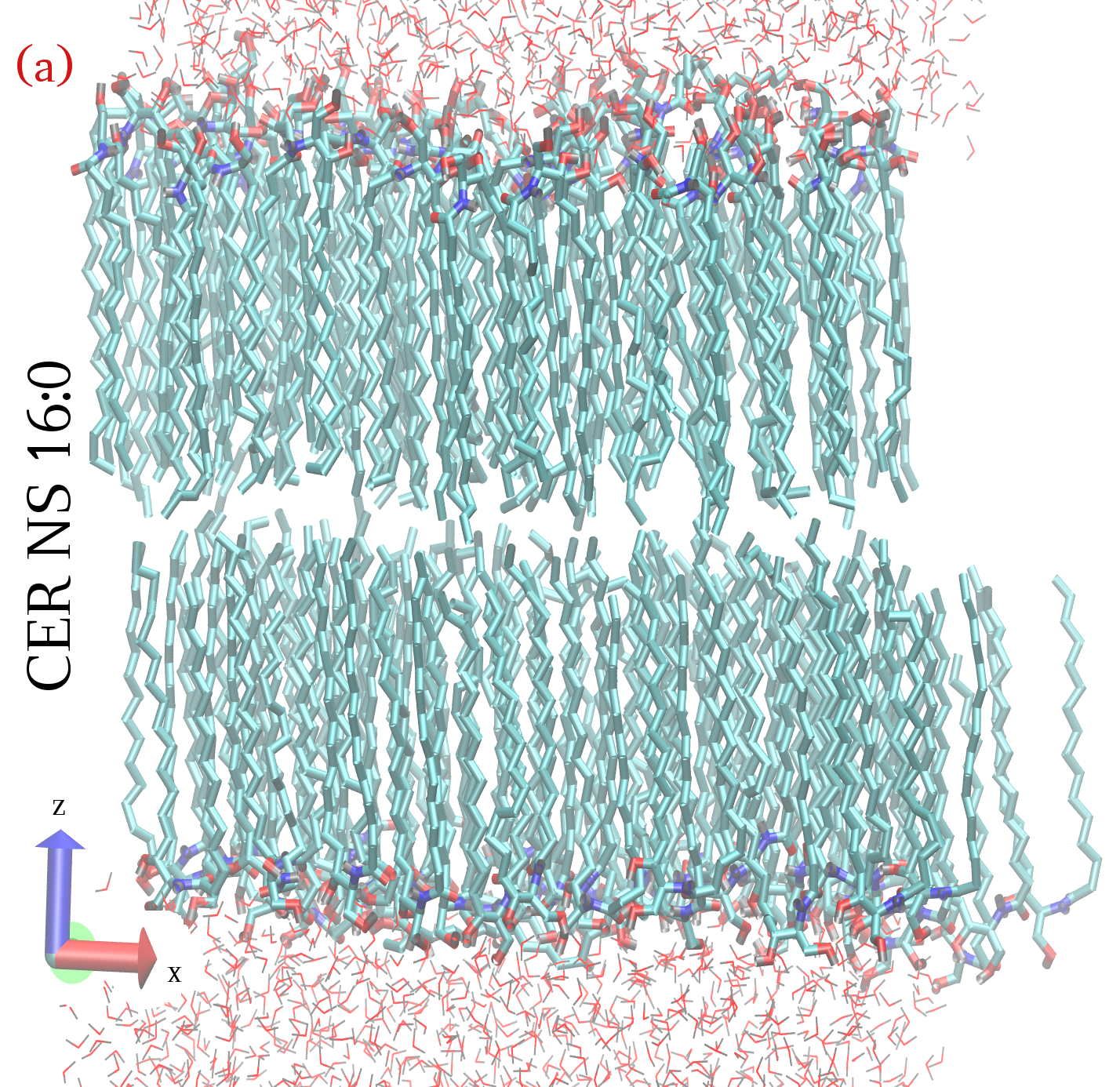} \hspace{0.02\linewidth}%
 \includegraphics[width=0.46\linewidth,clip=]{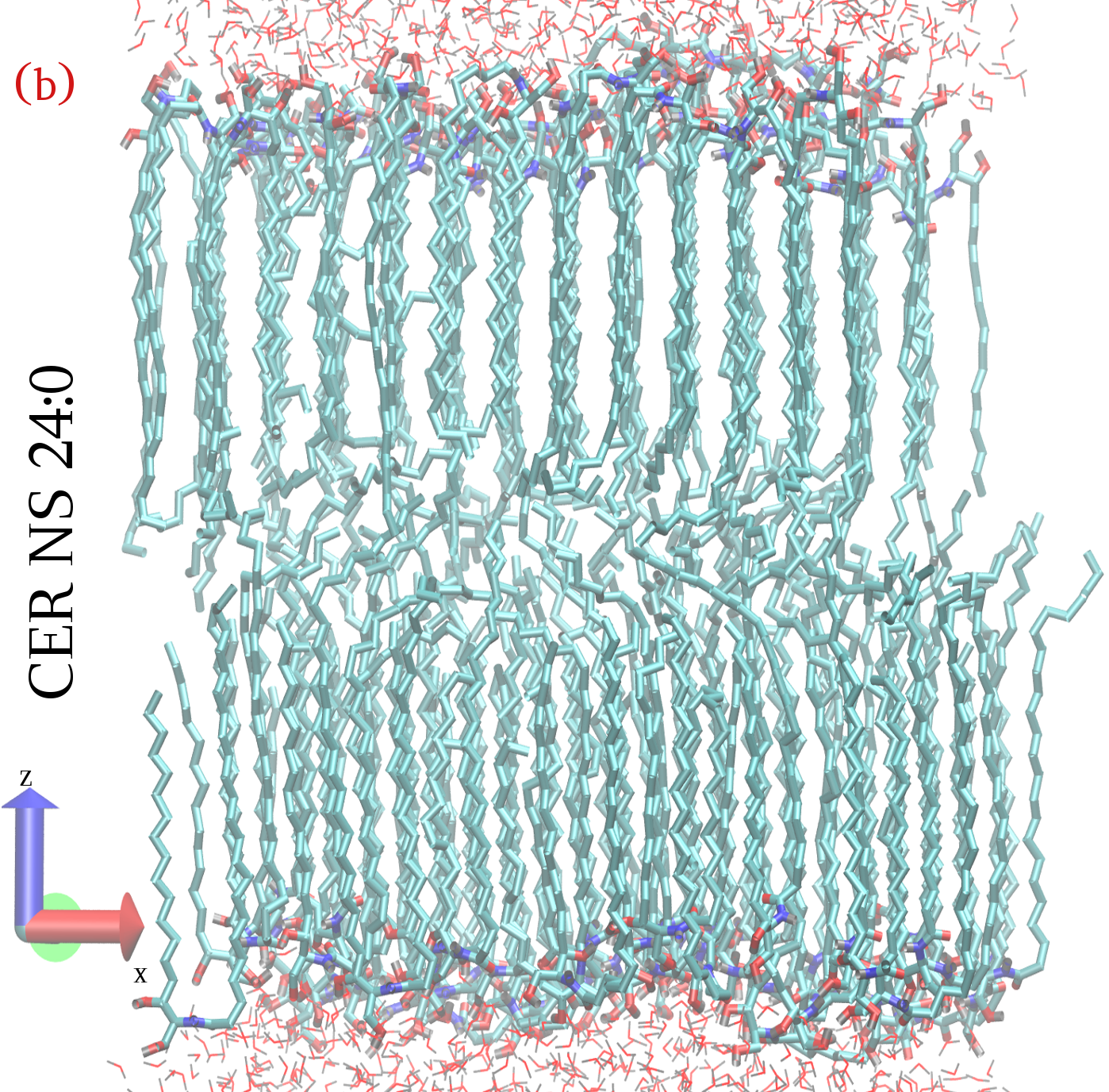} }
\centerline{ \includegraphics[width=0.46\linewidth,clip=]{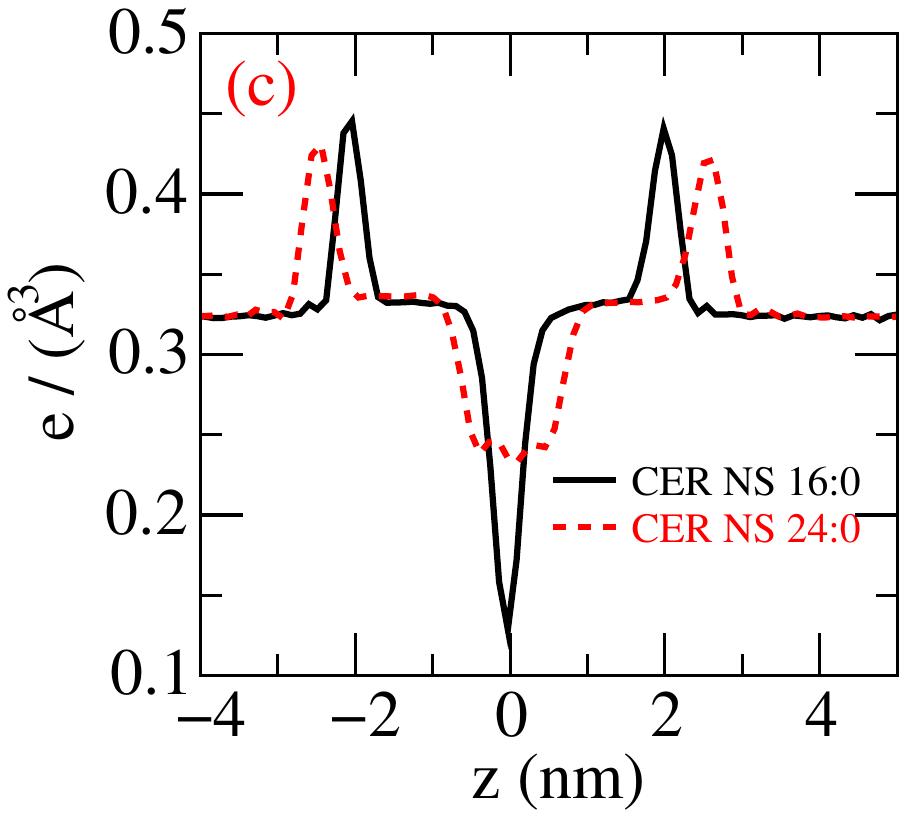} \hspace{0.02\linewidth}%
 \includegraphics[width=0.46\linewidth,clip=]{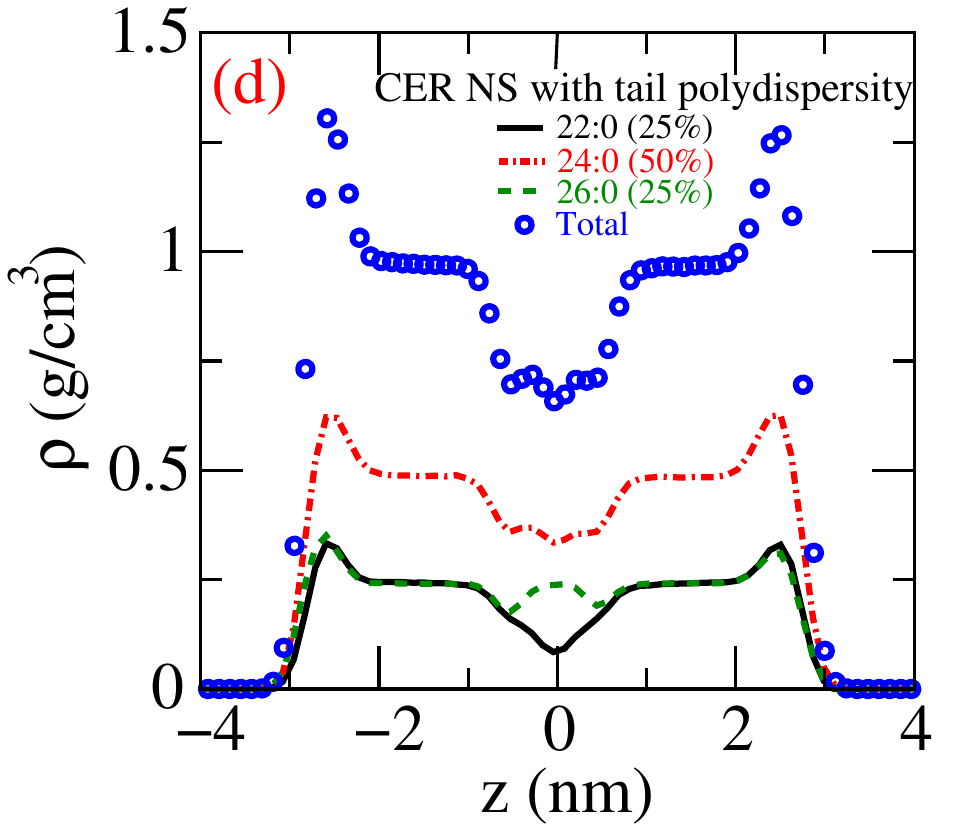} }
\caption{Molecular arrangements in (a) CER NS 16:0 and (b) CER NS 24:0 bilayers. The lipids are shown with thick lines. Only
a thin layer of water molecules near the lipid headgroups are shown with thin lines. (c) Electron density profiles for CER NS 16:0 (solid line)
and for CER NS 24:0 (dashed lines). The box dimensions for CER NS 16:0 were $L_x=5.43\,\textrm{nm}$, $L_y=4.66\,\textrm{nm}$,
and $L_z=11.19\,\textrm{nm}$. The box dimensions for CER NS 24:0 were $L_x=5.45\,\textrm{nm}$, $L_y=4.69\,\textrm{nm}$, and
$L_z=12.07\,\textrm{nm}$. (d) Densities of CER NS 22:0 (solid line), CER NS 24:0 (dot-dashed line), and
CER NS 26:0 (dashed line) in 1:2:1 mixture of the three lipids. The symbols show the total density.}
\label{fig.tailp}
\vspace*{-5pt}
\end{figure}

Fig.~\ref{fig.tailp} (a) and (b) respectively show typical molecular arrangements for CER NS 16:0 and for 
CER NS 24:0. CER NS 16:0 with symmetric tails shows a sharp inter-leaflet boundary and both of the tails remain 
perfectly ordered. In contrast, the ends of the longer fatty acid tails
of CER NS 24:0 assume a liquid-like disordered structure, giving the bilayer a sandwich ordered-disordered-ordered
structure \cite{das.bpj.09}. The longer fatty acid tail is further supported by dynamic bending at the head group region which gives 
rise to a slithering motion of the tails back and forth along the layer normal direction. 
The electron density of CER NS 16:0 in Fig.~\ref{fig.tailp} (c) shows a prominent dip in the bilayer midplane. 
There is a constant electron density region between the peak at the headgroup positions 
($z \simeq \pm 2\,\textrm{nm}$) and the bilayer midplane ($z=0$) determined by the length of the tails. For CER NS 24:0, 
the peaks at the head groups are further apart ($z \simeq \pm 2.5\,\textrm{nm}$) and broader, reflecting the slithering
motion of the tails. The liquid-like overlap region shows much larger electron density in the bilayer mid-plane than CER NS 16:0. The constant density
regions of both CER NS 16:0 and CER NS 24:0 bilayers have  the same density and width, reflecting the fact that the ordered region is 
determined by the sphingosine tails of
same length shared between the two species. Fig.~\ref{fig.tailp} (d) shows the lipid mass density (symbols) in a mixed 
bilayer comprising a 1:2:1 molar ratio of CER NS 22:0, 24:0, and 26:0. The densities from individual components show that 
the shorter tailed CER NS 22:0 (solid line) has a dip in the bilayer midplane, the longer tailed CER NS 26:0 has a peak (dashed line),
and CER NS 24:0 (dot-dashed lines) has comparatively flat profile. 
Thus, the bilayer thickness is determined by the average lipid  tail length, with shorter-tailed lipids having straighter conformations and longer-tailed lipids having
the excess carbon atoms disordered in a liquid-like conformation. 

\begin{figure}[htbp]
\vspace*{-7pt}
\leftline{\hspace{0.1\linewidth}CER NS 24:0 \hspace{0.2\linewidth} CER NP 24:0 \hspace{0.12\linewidth} CER NS + CER NP}
\centerline{%
 \includegraphics[width=0.32\linewidth,clip=]{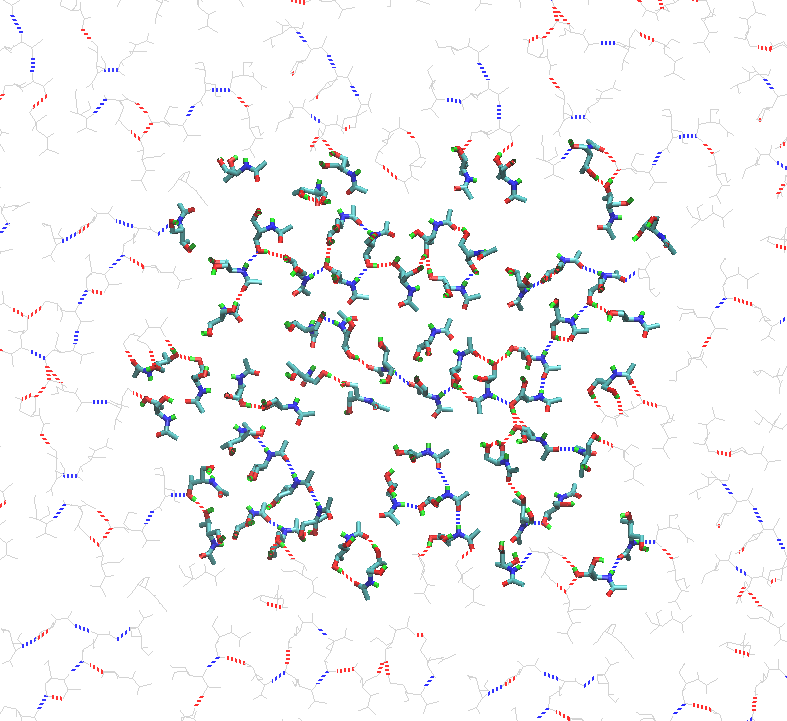} \hspace{0.02\linewidth}%
 \includegraphics[width=0.32\linewidth,clip=]{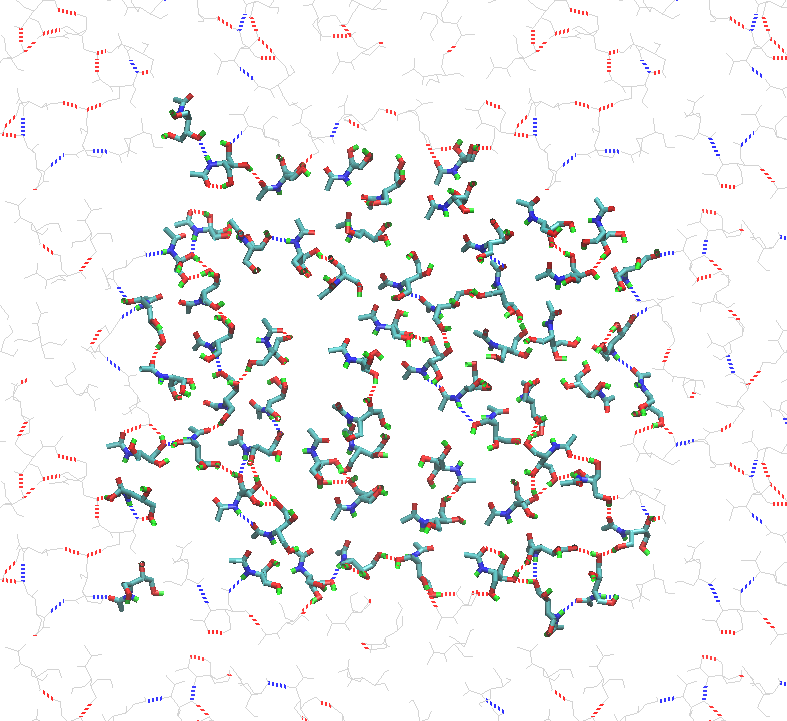} \hspace{0.02\linewidth}%
 \includegraphics[width=0.32\linewidth,clip=]{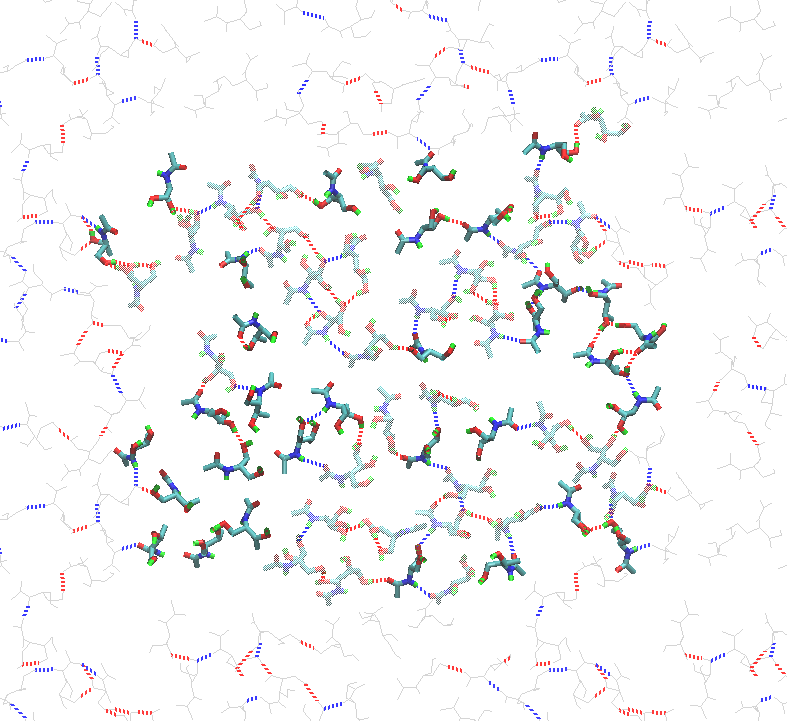}}
\vspace*{0.1cm}
\centerline{\includegraphics[width=0.32\linewidth,clip=]{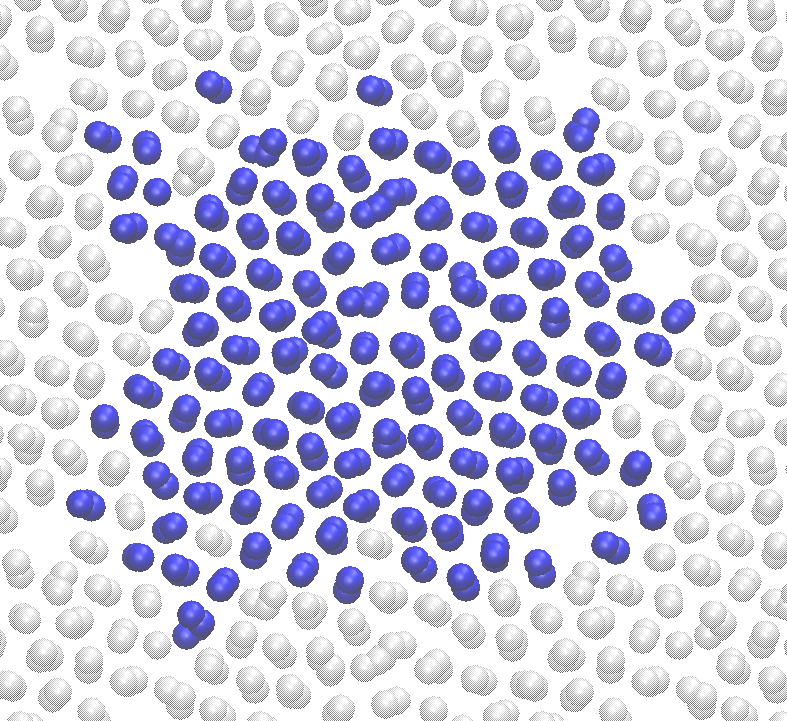} \hspace{0.02\linewidth}%
 \includegraphics[width=0.32\linewidth,clip=]{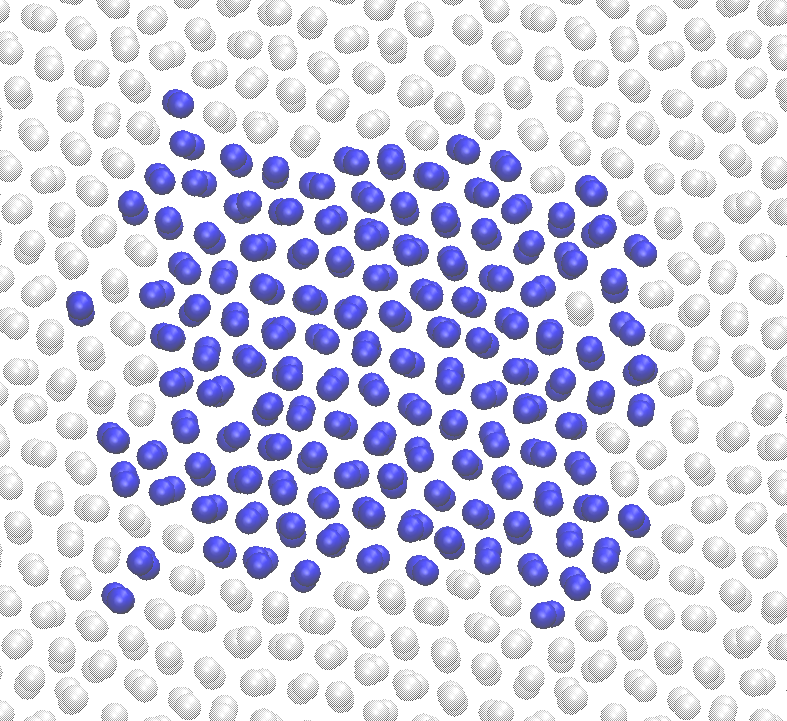} \hspace{0.02\linewidth}%
 \includegraphics[width=0.32\linewidth,clip=]{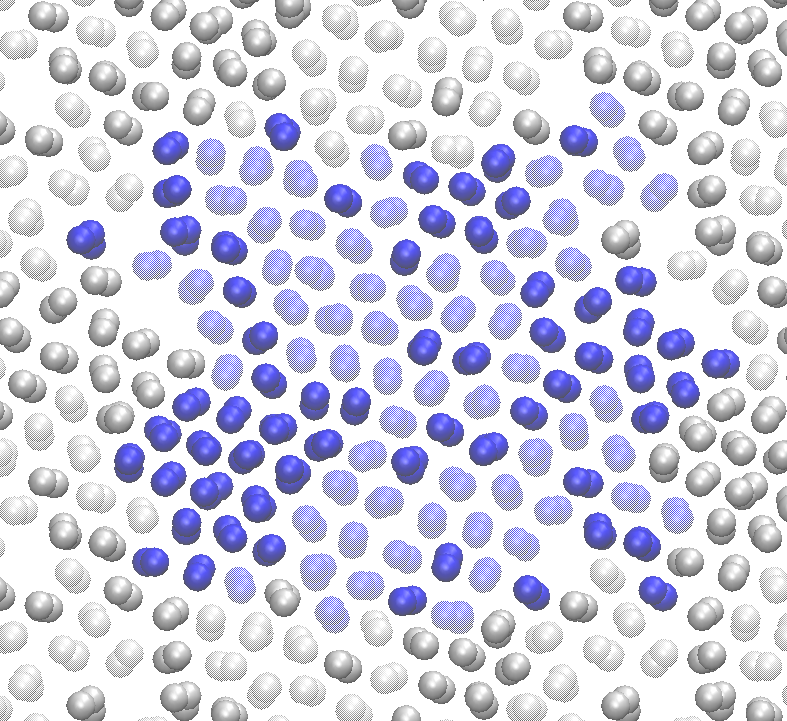}}
\caption{Snapshots from equilibrated simulations of three different bilayers (CER NS 24:0, CER NP 24:0 and an equimolar mixture of CER NS 24:0 + CER NP 24:0) showing the hydrogen bonding behaviour.  Top panel, Lipid hydrogen bond network:
Inter-lipid hydrogen bonds are shown as dashed lines (blue and red colors indicate hydrogen bonds
involving nitrogen and oxygen atoms respectively). Only the atoms in the head group region from the top leaflets are shown, and the 
molecules in the central simulation box shown as thick bonds. A small portion of the surrounding periodic images is included as thin gray lines to
highlight the connectivity of the inter-lipid hydrogen bond networks. In the right hand panel, CER NS are shown with solid bonds and  CER NP are shown as shaded bonds. 
Bottom panel, 2-d packing of tail carbon atoms: Only atoms in a $0.3\,\textrm{nm}$ slice
just below the headgroup region are shown. The atoms from the central simulation box are shown as blue spheres and a small portion of the surrounding periodic images is included in gray. In the right panel, atoms from CER NS are shown as solid spheres and those from
CER NP are shown as shaded spheres. The simulation box-sizes for the CER NS system were $L_x=5.45\,\textrm{nm} \times L_y=4.69\,\textrm{nm}$;
for the CER NP system $L_x=5.44\,\textrm{nm}\times L_y=4.74\,\textrm{nm}$; and for the mixed bilayer $L_x=5.40\,\textrm{nm}\times L_y=4.70\,\textrm{nm}$.}
\label{fig.hbnd}
\vspace*{-5pt}
\end{figure}

\subsection{Head group polydispersity and inter-lipid hydrogen bonds}

Fig.~\ref{fig.hbnd} compares the hydrogen bonds and tail packing in bilayers comprising CER NS 24:0, or CER NP 24:0, or an equimolar mixture of the two. 
CER NS has four sites that are capable of forming inter-lipid hydrogen bonds, while CER NP has one more acceptor-donor site.
In principle, either lipid can form a percolating a two-dimensional network, which  provides an elastic contribution to the mechanical response of the bilayers (such as the stretching modulus, or the bending moduli). In principle a percolating network of long-lived bonds assures a solid phase. However, 
the excluded volume of the alkyl tails prevents configurations
with such an infinite hydrogen-bond cluster because the average hydrogen-bond separation is smaller than the acyl tail separation. Instead, the head groups of a small number of lipids can tilt together in a cluster to form inter-lipid hydrogen bonds. 
The groups not involved in forming hydrogen bonds are energetically not penalised since
they share hydrogen bonds with water molecules. The clustering of the head groups has little effect on the
packing of the alkyl tails. The lower panels in Fig.~\ref{fig.hbnd}, of the methylene groups in a $0.3\,\textrm{nm}$ slice just below the headgroup region, 
show that the  tails form a local hexagonal arrangement.

\subsection{Conformation of CER EOS in a realistic SC lipid multilayer stack}

\begin{figure}[htbp]
\vspace*{-7pt}
\centerline{%
 \includegraphics[width=0.8\linewidth,clip=]{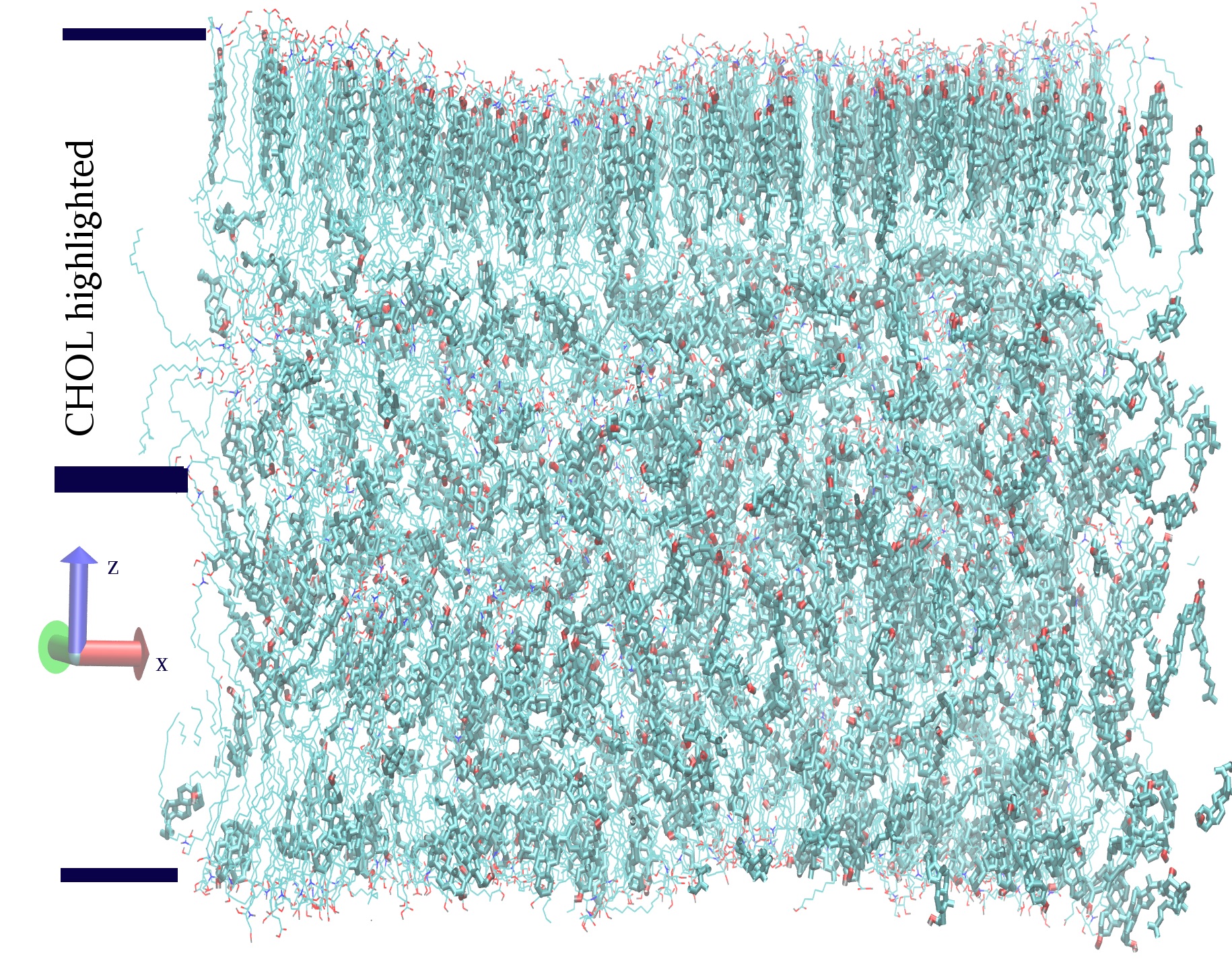} }
\centerline{%
 \includegraphics[width=0.8\linewidth,clip=]{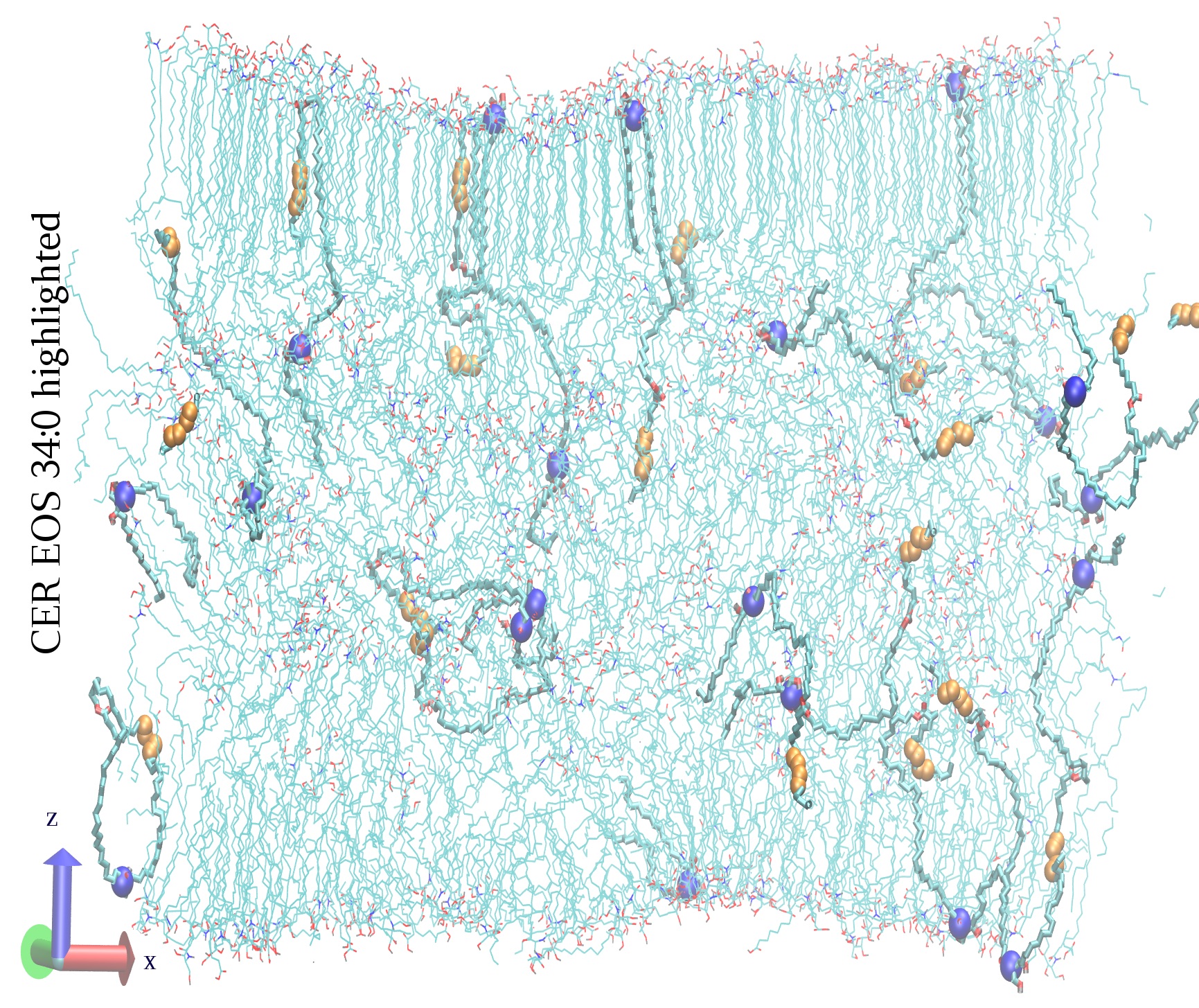} }
\caption{
Top panel: Final configuration from the simulation of multicomponent double bilayer. Only lipids in a $10\,\textrm{nm}$
slice along the $y$-direction are shown. The cholesterol molecules in this slice have been highlighted with thicker lines.
The water molecules surrounding the double bilayer have been omitted. The bars represent approximate
boundaries of the two bilayers. Bottom panel: The same simulation slice is shown omitting the CHOL molecules and highlighting 
CER EOS molecules with  N-chains  34 carbons long. In the highlighted CER EOS molecules, the head group position is
further highlighted by drawing the Nitrogen atom with large blue spheres. The position of the two cis-double bonds in the
conjugated linoleic acid is marked with orange spheres. Note that because we have plotted only a slice of the simulation box,
parts of some of the molecules are missing in the figure.}
\label{fig.mxd}
\vspace*{-5pt}
\end{figure}

Fig.~\ref{fig.mxd} shows the lipid arrangement at the end of $300\,\textrm{ns}$ simulation of the highly polydisperse
multilayer. For clarity, we have only shown atoms in a $10\,\textrm{nm}$ slice along the $y$-direction (into the page). 
In the top panel the bilayer boundaries are indicated  by  horizontal bars, and the  CHOL molecules have been highlighted. This structure 
(with roughly $32\times32$ lipids per leaflet)  is much more disordered than in the
small CER bilayer simulations presented in previous subsections and in similar sized bilayer simulations containing
CER NS 24:0, FFA 24:0, and CHOL \cite{das.bpj.09}.
This simulation is consistent with previous studies that identified
CHOL as a fluidiser for long tailed lipids \cite{kitson.biochem.94}. The chair-like rigid structure of CHOL disrupts tail order and, at high CHOL concentrations, 
even for simpler three component SC bilayers, some of the CHOL molecules reside in the
low-density inter-leaflet region \cite{das.bpj.09}. The distribution of CHOL among the ordered and disordered regions is dynamic, 
with comparatively fast exchange via flip-flop \cite{Das.sm.14}. In the bottom panel we omit the CHOL molecules
and highlight only those CER EOS molecules whose N-chain is 34 carbons long (excluding the conjugated
linoleic acid). The head group position for these molecules have been highlighted with blue spheres and the
unsaturations in the conjugated linoleic acid has been highlighted with smaller orange spheres. None of these CER EOS molecules
were found to simultaneously occupy both bilayers. Instead, the conjugated linoleic acid often folds close to the conjugation point, as well as at the unsaturated cis-double bonds. Occasionally some of the molecules were found
to have the conjugation point (with partial changes and hydrogen bond forming capability) at the bilayer boundary and 
the head group buried inside the lipid interior, either forming hydrogen bonds with inter-leaflet CHOL or dragging some fatty acids into the bilayer interior. 

\section{Discussion} 

\subsection{Elastic Properties and Membrane Morphologies}
While the results presented in this paper investigate properties of SC lipids in a lamellar structure, it is far from clear whether such a structure is
thermodynamically stable. Integrals of moments of the excess lateral pressure profile along the bilayer normal can be exploited to estimate the curvature
moduli of the bilayer and the intrinsic curvature preferred by the leaflets.

In earlier simulations the excess lateral pressure profiles $\delta P
(z)=\tfrac12\left[p_{xx}(z)+p_{yy}(z) \right] - p_{zz}(z)$ were calculated \cite{das.bpj.09} for 
bilayers with different molar ratios of CER NS 24:0, CHOL, and FFA 24:0 using the same force field parameters as in this paper and
with a group-based cut-off for the electrostatics. 
Here $p_{ij}(z)$ is the local pressure tensor. The excess lateral pressure profile $\delta P$ is believed to control the behaviour of membrane proteins in 
phospholipid membranes \cite{Cantor97}. The integral of moments $z^{\alpha}$ of $\delta P$ 
leads to curvature elastic constants \cite{schofield.localp.82,goetz1998computer}:
\begin{align}
\kappa_{\scriptscriptstyle M} C_{0} &= -\int_0^{d/2}dz \,z\, \delta P(z) \label{eq:c0}\\
\bar{\kappa}  &= \int_{-d/2}^{d/2}dz\, z^{2}\, \delta P(z), \label{eq:kappabar}
\end{align}
where $z = 0$ is at the bilayer midplane,
and $d/2$ is sufficiently large such that membrane stresses vanish (and will typically be 
several water layers larger than the thickness of a single leaflet of the bilayer). Here, $\kappa_{\scriptscriptstyle M}$ is the mean curvature modulus of one monolayer
(often assumed to be half of the bilayer mean curvature modulus $\kappa$), 
$C_{0}$ is the spontaneous curvature of the monolayer, and $\bar{\kappa}$ is the Gaussian curvature modulus of the bilayer. Estimates of
$\kappa$, $\bar{\kappa}$, and $C_0$ for selected bilayers are shown in Table \ref{tab.kappa}.
\begin{table}[!h]
\caption{Gaussian Curvature moduli estimated using Eq.~\ref{eq:kappabar} from simulations \cite{das.bpj.09} of CER 24:0, CHOL and FFA 24:0 at $T=340\,\textrm{K}$ (\cite{das.bpj.09} did not calculate these properties explicitly).}
\label{tab:moduli}
\begin{center}
\begin{tabular}{cllll}
\hline
Composition &${\kappa}/\kbT$  &$\bar{\kappa}/\kbT$ & $\bar{\kappa}/\kappa$ & $1/C_0$ \\
(CER:CHOL:FFA) & & & & (nm) \\
\hline
1:0:0 & 140 & 91 & 0.7 & -50\\
1:1:0 & 98 & 90 &0.9 & -35 \\
1:0:1 & 75 & 82 & 1.1 & -32 \\
1:1:1 & 117 & 75 & 0.6 & -54 \\
2:2:1 & 83 & 91 & 1.1 & -35 
\end{tabular}
\end{center}
\label{tab.kappa}
\vspace*{-4pt}
\end{table}

The free energy per area $f$ of the membrane due to bending is given by 
\begin{equation}
f = \tfrac12\kappa(c_{1}+c_{2}-C_{0})^{2} + \bar{\kappa}c_{1}c_{2},
\end{equation}
where $c_{1}$ and $c_{2}$ are the (local) principal curvatures of the membrane.
For $\bar{\kappa}<0$ a membrane prefers spherically curved shapes, rather than saddle shapes, while for $\bar{\kappa}>0$ a membrane is unstable to saddle-like shapes, 
characteristic of bicontinuous phases \cite{templer1998gaussian}.

Deserno and co-workers \cite{hu2012determining,hu2013gaussian} have studied the curvature moduli obtained using this method in coarse-grained simulations based on the MARTINI force-field. They concluded that this method may provide inaccurate results because the moment calculations may neglect important correlations in the bilayer \cite{C2FD90043B}. With this caveat in mind, we nonetheless report our calculations of the Gaussian curvature modulus $\bar{\kappa}$, noting that we expect the more atomistic approach taken here to be more accurate than coarse-grained calculations (this point was also studied in \cite{C2FD90043B}).
For the SC lipid membranes we find $\bar{\kappa}\sim+75-90\kbT$, which signifies that an isolated membrane is unstable towards forming saddle-shaped surfaces. We find the following main features:
\begin{enumerate}
\item  The value for $\bar{\kappa}$ is large; while we cannot yet measure the mean curvature modulus directly from the simulations \cite{hu2012determining}, we can
crudely estimate $\kappa$ by calculating the area compressiblity modulus $\kappa_A$ from the area fluctuation and connecting $\kappa_A$ and the bilayer thickness
to $\kappa$ with the polymer brush theory \cite{Rawicz2000}. The calculated values of $\kappa$ for SC lipid bilayers are $\sim 75-140 \kbT$, which like most membranes is of the same order as $\bar{\kappa}$ \cite{hu2012determining,hu2013gaussian}.
\item A large positive value for $\bar{\kappa}$ would stabilize bicontinuous lamellar phase, which is consistent with cryo-EM images showing a cubic structure in the interstices of the corneocytes \cite{Amoudi.JID.05} of \textit{ex vivo} sections.
 \item Simulations on moderately large  systems ($\simeq 32 \times 32$ lipids per leaflet, 4004 lipids in total) \cite{das2013lamellar} demonstrated that a hydrated double bilayer is unstable to a cylindrical shape. However, the curved state was cylindrical with Gaussian curvature $c_{1}c_{2}=0$, which is the same as for the lamellar state and not sensitive to $\bar{\kappa}$. It is possible that the finite system size led to trapping in a cylindrical state on the way to a saddle shape that was incommensurate with the size of the simulation.
\item The inverse micellar state found in \cite{das2013lamellar} upon compressing a dilute gas of water and lipids to room temperature and pressure is consistent with $\bar{\kappa}<0$ rather than $\bar{\kappa}>0$. 
\end{enumerate}
These results for the bending moduli, combined with the observations of a non-lamellar structure in the native SC \cite{Amoudi.JID.05}, suggest that further systematic study of the detailed lipid morphology in model and native SC lipid mixtures is warranted.

\subsection{Summary}
In this paper we have reported results from atomistic simulations of bilayer and multilayer arrangements of a series of different combinations of SC lipids
to isolate the effects of different kinds of polydispersity seen in skin lipids. Both the asymmetry and tail polydispersity was found to enhance 
partial tail-interdigitation (a region occupied by chains from both the leaflets). The degree of interdigitation is closely associated with inter-leaflet friction \cite{denOtter.BPJ.07}. 

Somewhat surprisingly, we found that the number of hydrogen bond-forming groups on the headgroup of ceramides play little or no role in the 
size of hydrogen-bonded lipid clusters. This is because the separation of a hydrogen bond is smaller than the  densest possible tail packing, which leads to small cluster sizes. 

The very long-tailed asymmetric lipid CER EOS was found to remain in a single bilayer and reduce the tail order substantially. Previous simulations with multicomponent 
SC lipids that contained a branched CER EOS (where linoleic acid was replaced with a saturated fatty acid 
containing a methyl group at the place of the unsaturation) found that, from an initial straight tail arrangement, 
 the majority of CER EOS molecules remained embedded in
the next bilayer \cite{Engelbrecht.smat.11}. The differing findings may be due to different protocols adopted in forming the
initial structure. In our simulations, although CER EOS was given an initial straight hairpin initial conformation, 
the starting configuration comprised molecules placed in a loosely packed structure. Hence, CER EOS 
had the freedom to fold back and remain in the same bilayer.
Because of the reduced mobility in multi-lamellar structures of SC lipids, a definitive answer to the equilibrium conformation adopted by CER EOS would require an expensive free-energy calculation that has not been attempted here.

However, equilibration may in fact not be relevant for understanding the SC \textit{in vivo}. The much higher than physiological temperature for chain-melting of SC lipids and strong confinement between corneocytes may make the
pathway of formation far more important for SC lipids than for liquid disordered bilayers of phospholipids. Ceramides are converted to galactosylceramide by
addition of a sugar moeity immediately after synthesis in the stratum granulosum. Simulations\cite{Hall.PCB.11} show that galactosylceramides occupy an
area/lipid similar to  sphingomyelin in bilayers containing sphingomyelin and CHOL, which is much larger than the 
typical area/molecule in the SC membranes. 
The presence of this large sugar moeity will disfavor the extended tail
arrangement  seen in some CER crystal structures and hypothesized in some models of SC lipid arrangement \cite{iwai.jid.12},   and make hydrated bilayers
thermodynamically stable. Only once the unilamellar vesicles have been transported out of the cells 
is the sugar moeity  removed, which then can lead to a gel-like membrane; at this point 
 the vesicles seem to lose internal water and fold into  a stacked disc-like structure \cite{Madison.JID.87}.  
If the lipids are already in gel-phase by the time the SC lipid vesicles are flattened and stacked, then it seems likely that the long CER EOS will remain confined in bilayers.

There have been surprisingly few simulations of SC bilayer phases \cite{notman.dmso.07,palycyova2015,wan2015interaction,gupta2015molecular}. There is ample room and numerous problems to study as simulation power grows. The interaction of the SC matrix with corneocytes is a fascinating question, as is the accompanying evolution of the lipid structure from living cells in the lower epidermis to the fully formed SC barrier. The effects and properties of additives is of obvious importance for regulating and treating skin and understanding disease, and the mechanical properties of the full SC layer, incorporating the mechanics of both corneocytes and lipid bilayers, remains a long way off. More detailed experiments that combine mechanical or chemical stress with morphology, dynamics, and structure, ideally at the 5-100 nm scale are essential in order to provide benchmarks and fully understand the remarkable resilience of the SC. 

\ack{We thank Massimo Noro (Unilever Research) for inspiration, encouragement, and discussions; and Anna Akinshina and John Seddon for helpful discussion.}

\funding{This work was supported by Yorkshire Forward (YFRID Award B/302) and Unilever Research, and part financed by the European Regional Development Fund (ERDF). This work made use of the facilities of N8 HPC Centre of Excellence, provided and funded by N8 consortium and EPSRC (Grant No. EP/K000225/1). Additional
computing resources were provided by SoftComp EU Network of Excellence.}

\conflict{We have no competing interests.}

\contributions{CD conceived of and designed the study, carried out the simulations, performed the analysis, and co-wrote the manuscript. PDO conceived of the study and co-wrote the manuscript.}

\dataccess{The force fields, molecular topology and the final configurations 
for the simulations presented here are available at 
\url{http://dx.doi.org/10.5281/zenodo.49270}.}

\bibliographystyle{vancouver}   
\bibliography{RSBib}
\end{document}